%% file: agents.tex
\documentclass[11pt]{article}
\usepackage{geometry}                
\usepackage{graphicx}
\usepackage{amssymb}
\usepackage{epstopdf}
\usepackage{float}
\restylefloat{table}

\input{epsf}

\title{On Agents and Equilibria}
\author{Ted Theodosopoulos\footnote{Department of Mathematics, Saint Ann's school, Brooklyn, NY, USA}}
\date{}                                           

\begin{document}
\maketitle

Ever since my economics class debated the creeping hegemony of market mentality in human affairs last spring I have wanted to articulate the surprise and wonder I felt at my students' probing of {\it homo economicus} and his boundaries\footnote{A summary of the points debated and the readings that informed them can be found at https://sites.google.com/a/saintannsny.org/saint-ann-s-math-team/economics-pages/economics-debate-2012-13.}.  Later in the summer, my encounters at a conference on new approaches to economics\footnote{https://sites.google.com/site/wehia2013/.} strengthened my conviction of the utility of such an essay.  I have finally gathered my thoughts sufficiently to put them down.

What is {\it homo economicus}, but a conception of the economy as a sphere of individual agents, acting in a rational pursuit of maxima of their endowed utilities, and interacting by constraining each other's, and their own, available options, due to fundamental resource limitations.  It is a mindset that seeks to identify alternative actions in every instance, and enunciate a tradeoff that encompasses the decision-making process of the agent at hand.

It would be useful to put forth a different narrative of the economic sphere, if only as a check to the range of applicability for the dominant paradigm.  It is, after all, at the interface between competing conceptualizations that elements of reality seep through.  In this spirit, can we lift the conceptual burden of decades, convincing us that conflicting alternatives are the sole rational view of reality, and construct a plausible paradigm that evades this pervasive market mentality?

My students came up with the notion of {\it homo socialis}, a collection of agents that seek to maximize their social interactions, their ÒconnectednessÓ, as it were, without regard to the potential economic costs and benefits entailed in this pursuit\footnote{This shares much with Helbing's writings on this subject, for example http://www.nature.com/srep/2013/130319/srep01480/pdf/srep01480.pdf.}.  Perhaps, after all, utility isn't one-dimensional and its components aren't exchangeable.  Moreover, maybe utility isn't assignable to individuals, but is instead an attribute of agent configurations encoding their state as well as their interactions.

Envisioning such apparently incongruous possibilities is the hallmark of a revolution that is quietly engulfing economic theorizing.  And it isn't primarily the recent cascade of global crises that fuels it, but rather the shifting aesthetic of what constitutes convincing narratives in our increasingly interdependent information web.  The stalwart, mechanistic paradigm, which sought to confer impartial objectivity and monopolistic conceptual authority to disembodied aggregate forces harks back to the Enlightenment image of the clockwork physical universe, that defies the need for spontaneous creativity, hopes and imaginings.  It was a world of consistent order, a cosmic billiards in which collisions conferred indubitable truth, and bounces were eternally predictable, if only we possessed sufficient intellectual capacity.  As if the physical, not to mention the economic, realm were there all along, apart from us, and all we needed to do is accept its incontrovertible workings$\ldots$

Our models of the physical universe have moved stupendously since then, but a vast edifice was built for understanding the economy based on this long-surpassed model.  Physicists learned that objects and actions coexist in an inseparable reciprocal nexus of transformations, where reality attains coherence through mutually reinforcing potential configurations living beyond space, time and incongruity.  Mathematicians broadened their perspectives immensely to accept the inherent, perpetual incompleteness of the quest for incontrovertible knowledge, and learned to thrive in this newly open logical paradigm.  It is time that our notion of the economic sphere evolved beyond the centuries-old set of incompatible competing alternatives that still suffuse it.

There are two concrete re-conceptualizations I'd like to elaborate on here, in hopes of clarifying the path I am striving to describe, and to illustrate convincingly the conceptual burden I alluded to earlier.  The first one seeks to debunk the seemingly self-evident necessity of populating reality with mutually incompatible alternatives, thereby representing the space of choices as a geometric space of points\footnote{Technically, our objection is to the standard topology that is taken for granted on this `choice space'.  After all, it is always possible to represent any set as a set of points, albeit with a decidedly nontrivial topology, to appropriately reflect the nature of nearness that is natural for the original set.}.  Instead, I propose that we envision choices as configurations of a network, irreducible to actions at specific nodes.  For several years now, Brandenburger and others have emphasized the irreducible relevance of contextuality in our decision-making, more akin to evolving probability distributions over our joint actions, in a manner resembling the entangled evolution of quantum states\footnote{See for example http://iopscience.iop.org/1367-2630/13/11/113036/.}.

After all, the value I assign to various potential actions available to me, and even my conception of this set of available actions, depends inextricably on my appreciation of actions by my neighboring agents, and my expectations of their appraisal of my actions, not to mention the very set of agents I currently consider my neighbors.  At any point in time the joint specification of these variables for all the agents is almost certainly strewn with inconsistencies, rendering the vision of a Òrepresentative agentÓ worse than chimeric, as Kirman has argued for two decades now\footnote{http://www2.econ.iastate.edu/tesfatsi/WhomOrWhatDoesRepIndRepresent.AKirman1992.pdf.}.

We navigate these puzzling, incompatible, probabilistic assessments with a scant thought about the theoretical conundrums they cultivate.  Our choices aren't driven by a desire to find incontrovertible truth.  Instead, we construct narratives to convince ourselves and our neighbors of the logic of our actions, unbothered by the occasional necessity of shifting perspective.  In fact, our success in navigating the intricately intertwined paths that criss-cross the economic sphere rests largely on this uniquely creative ability to re-invent our reasoning, introducing new, unforeseen twists and wrinkles to what appeared a smooth explanation earlier.

Such re-imaginings of our surroundings, both actual and potential, cannot be accessed through an evolution on finite-dimensional spaces of alternatives.  We need instead to consider infinite-dimensional path spaces, encompassing historical narratives that go beyond any finite listing of their turns.  We need to accommodate evolving probability distributions over the space of counterfactual configurations of our information nexus, because many possible paths contribute simultaneously to the creative reasoning behind our actions.

This first recasting of our economic vision updates the agents that populate the economy, from distinct individuals, to hierarchically organized objects, with components of differing dimensionalities, fitting together in geometrically intricate, irreducible evolving patterns.  The second re-imagining I propose involves the questions we ask of these hierarchical, meta-individual agents.  What patterns characterize their observable behaviours?  Do they give rise to the typical structures and institutions that we have come to expect?

In this instance we have to confront another deeply held, if unquestioned, conviction, which turns out to be less justified, often misleading us to comforting misapprehensions.  The notion I am after involves the interplay we take for granted between time and randomness.  What do we mean when we say that an event, like an even number in a die toss, occurs with some probability, say ${\frac {1}{2}}$?  Strictly speaking, we mean that if we could lay out and count all the possible outcomes of an experiment, half of them will have the desired property.  But often it's impractical to literally survey all possible outcomes, even when they aren't infinite.  After all, the overwhelming majority of numbers are forever beyond our direct access, even though they are, by definition, finite.  What do we do in lieu of this na\"ive counting?

In order to probe this question deeper, imagine confronting the question of fairness for the die.  How would you distinguish a biased die from a fair one?  You'd roll it many times and compare the actual, what we call `empirical', outcomes to those expected under the hypothesis that the die was indeed fair.  If, for instance, we roll the die $N=100$ times and we obtain an even number $58$ times, we'd probably judge this event to be consistent with the hypothesis that the die was fair, which would have predicted ${\frac {1}{2}}$ of the outcomes, i.e. $50$, to be even.  In fact, statistics tells us that, under the `null' hypothesis of a fair die, the number of even outcomes out of $N=100$ tosses is itself random, with `expected value' equal to ${\frac {N}{2}}$, i.e. $50$, and `standard deviation' equal to $\sqrt{N \left({\frac {1}{2}} \right) \left(1 - {\frac {1}{2}} \right)}$, i.e. $5$.  This latter concept is a measure of the expected spread of the randomness in the number of even outcomes out of $N=100$ tosses, assuming the coin was fair.  The actual measurement of $58$ even outcomes is only $1.6$ standard deviations above the expected value, a level which, statistics tells us, will not be exceeded with probability $0.9452$!

Let's marvel for a moment at the fact that, in order to make concrete sense out of a probabilistic statement, we wound around to another, more sophisticated probabilistic statement.  While this isn't actually circular thinking, and we have in fact moved to a higher rung in this conceptual helix, it is well worth pondering this logical `closure', or self-referential nature of probability.  But this isn't our present quest.  And neither is the remarkable fact that, our reasoning led us actually to doubt the fairness of our coin with more than 5\% confidence!  In other words, after this observation, it would be rational, even profitable, to bet $1:20$ odds that the coin was biased!

No, our quarry lies deeper.  In order to catch a glimpse, imagine that our $N=100$ tosses played out in one of the following three ways:
\begin{enumerate}
\item First we experienced $58$ even rolls and then $42$ odd ones.
\item First we experienced $42$ even rolls, followed by $42$ odd ones, and finally another $16$ even ones.
\item We experienced a succession of $42$ even-odd roll pairs, followed by $16$ even rolls.
\end{enumerate}
The point of course is that, while all these outcomes look indistinguishable in the end, each would have led us to very different conclusions along the way.  For example, under the third scenario we would never have any reason to doubt the fairness of the die until after the first $84$ tosses, at which point we begin experiencing the pronounced `run' of $16$ even rolls.  Already after $10$ consecutive even rolls we should have grave doubts as to the fairness of the die because such a run will occur with probability less than 0.1\% if the die were fair.

But what of all the `balanced' tosses that preceded this one-sided run?  Should they not serve to sway our judgement? After all, we've observed more than $8$ independent sequences of $10$ tosses in a row, all of which were entirely consistent with the fair die hypothesis.  How are we to balance the accumulated evidence in favour of the `null' hypothesis with the more recent evidence against it?

What if the `fairness' of the die was not determinable, because it varied!  Imagine, for example, that after each roll, we added a dot to the hidden face of the die.  Remembering that a standard $6$-sided die has odd and even numbers opposite one another, we see that after every roll, we would be `conditioning' the next roll to have the same parity as the previous one!

While the analysis of this nonstandard die is nontrivial, it should be clear that the a priori symmetric rules and starting configuration must guarantee a symmetric probability distribution for the outcome over any number of rolls.  But, at the same time, it's equally clear that we are much more likely to experience unbalanced paths, with overwhelming majority of one parity or the other over time.  To illustrate this dichotomy more concretely, consider what happens over three consecutive rolls of a standard die versus this peculiar one we've just constructed, as shown in the table below:
\begin{table}[H]
\begin{tabular}{|c|c|c|} \hline
Event & $\begin{array}{l} \mbox{Standard die} \\ \mbox{probability} \end{array}$ & $\begin{array}{l} \mbox{Nonstandard die} \\ \mbox{probability} \end{array}$  \\ \hline
EEE & $1/8$ & $7/27$ \\ \hline
EEO & $1/8$ & $2/27$ \\ \hline
EOE & $1/8$ & $1/12$ \\ \hline
EOO & $1/8$ & $1/12$ \\ \hline
OEE & $1/8$ & $1/12$ \\ \hline
OEO & $1/8$ & $1/12$ \\ \hline
OOE & $1/8$ & $2/27$ \\ \hline
OOO & $1/8$ & $7/27$ \\ \hline
\end{tabular}
\end{table}

We see that, while the overall probabilities remain symmetric and balanced, the unbalanced paths become increasingly more likely.  For example, if we saw a run of three same parity rolls in a row, under the standard die scenario we could reject the hypothesis that the die was fair expecting to be wrong with probability ${\frac {1}{4}}$, while the same inference under the nonstandard die scenario would lead to a mistake more than half of the time.

The phenomenon at play here is what we call `ergodicity', or lack thereof, and this die model we are considering allows us to explore it in greater depth.  The standard die scenario is ergodic, because no mater what state of imbalance between odd and even tosses you happen to find yourself in, there is a sequence of plausible, if highly unlikely, rolls that will get you to any other state of imbalance you desire.  For example, following a run of $3$ even tosses in a row, you may experience with probability ${\frac {1}{128}}$ a run of $7$ odd tosses, bringing the imbalance, measured as the proportion of even tosses, down to $0.3$, even though the die was really fair.

On the other hand, the nonstandard die scenario is non-ergodic, because there are imbalance states from which you cannot reach all other imbalance states.  Imagine for example 2 even tosses in a row in such a way that two opposite pairs of faces of the die are both even, while the third pair of faces remains odd on one side and even on the other.  From this state of imbalance, three outcomes are still possible:
\begin{enumerate}
\item With probability ${\frac {2}{3}}$ one of the four, pairwise matched even faces comes up, and nothing changes.
\item With probability ${\frac {1}{6}}$ the sole remaining odd face comes up, changing the opposite side from even to odd, and therefore increasing the chance that the next toss will be odd to ${\frac {1}{3}}$.  Note that this die state will no longer change, and so we are, from now on, in a situation equivalent to that of a standard, but unfair die, with even probability ${\frac {2}{3}}$ and odd probability ${\frac {1}{3}}$.  Thus, the standard statistical argument we saw earlier tells us that, after $N=100$ more tosses, the expected number of even outcomes will be ${\frac {2N}{3}} = 66.\bar{6}$  with standard deviation equal to $\sqrt{N \left({\frac {2}{3}} \right) \left(1 - {\frac {2}{3}} \right)} = 4.\bar{6}$.

In other words, under this scenario, we would expect to see roughly twice as many even as odd tosses in the long run, even though the process is completely symmetric, and therefore fair, to begin with.  After all, with equal probability, we could have arrived at a situation where we would expect twice as many odd as even tosses in the long run.  In any case, either of these two situations would lead us to expect a $2:1$ imbalance over time, as we experience more and more tosses, but the a priori probability of odd vs. even tosses is clearly $1:1$, as the system is unbiased, so the probability of an even toss at any time, after many repetitions of the same experiment from scratch, is ${\frac {1}{2}}$.
\item With probability ${\frac {1}{6}}$ the remaining unmatched even face comes up, switching the last odd face to even, and making it impossible to get an odd number in any subsequent tosses.  Thus, the system is henceforth locked in an all-even monopoly.  Of course, with equal likelihood we could have arrived at a similar all-odd monopoly, in a different run of the same experiment.  Once again we see that, while the prior probability of obtaining an even number at any point is ${\frac {1}{2}}$, the empirical proportion of even tosses over time along any particular path of this experiment will be unbalanced.
\end{enumerate}

It isn't hard to imagine economic situations that behave this way.  In fact, we often act under the tacit assumption that our actions, and those of the other economic actors, will change the chances of experiencing different outcomes.  More to the point, we experience the economy as a historically contingent process, inextricably dependent on the path we happened to follow along the way to our current state, much like the diversity that populates the living world we inhabit.  This marked divergence between the a priori expectation over counterfactual alternatives and the a posteriori assessment of contingent outcomes is absent from the dominant economic paradigm, whose epistemology rests instead on the presumed inexorable discovery of a pre-existing, if unknown, equilibrium, through an impartial, mechanical balance of impersonal forces.  A far cry from our decidedly strategic, unbalanced, intricately temporal economic experience!

The emerging agent-based methodology that is making inroads among the economic literature in the past decade aims to explicitly incorporate these, and other qualitative deviations from the efficiently mechanistic view of Smith's `invisible hand'.  It offers many tantalizingly open research directions, incorporating the hard-won conceptual battles of physicists and mathematicians over the last century, not to mention hopes for overcoming the all-too-apparent policy limitations of the dominant paradigm in the face of ever-intensifying global crises.  It is my hope that this new thinking will help economic science emerge into an era of renewed conceptual prosperity and practical relevance for humanity's demanding coordination problems.

\end{document}